\documentclass[prl,aps,twocolumn,
]{revtex4-2}

\usepackage{graphicx}
\usepackage{dcolumn}
\usepackage{bm}
\usepackage{xcolor}
\usepackage{slashed}
\usepackage{float}
\usepackage{comment}
\usepackage{amsmath}
\usepackage[normalem]{ulem}

\newcommand{\apjl}{Astrophys. J. Lett.}

\newcommand{\aap}{Astron. Astrophys.}
\newcommand{\mnras}{Mon. Not. Roy. Astron. Soc.}

\newcommand{\nphysa}{Nucl. Phys. A}
\newcommand{\physrep}{Phys. Rept.}

\usepackage[colorlinks=True, citecolor=blue,linkcolor=blue, urlcolor=blue]{hyperref}

\newcommand{\nn}{\nonumber}
\begin{document}

\title{Neutrino Emissivity in $\Delta$-direct-Urca Processes Inside Neutron Stars}

\author{Yogeesh N}
 \email{p23ph0018@iitj.ac.in}
\affiliation{Department of Physics, Indian Institute of Technology Jodhpur, Rajasthan, India}
\author{Monika Sinha}
  \email{ms@iitj.ac.in}
 \affiliation{Department of Physics, Indian Institute of Technology Jodhpur, Rajasthan, India}

\author{Vishal Parmar}
\email{vishal.parmar@pi.infn.it}
\affiliation{INFN, Sezione di Pisa, Largo Bruno Pontecorvo 3, I-56127 Pisa, Italy}

\date{\today}

\begin{abstract}
Observed massive neutron stars may contain heavier baryons such as $\Delta$ isobars and hyperons in their interiors. In the presence of $\Delta$ isobars, the neutrino emitting nucleonic Urca processes can be accompanied by $\Delta$-direct-Urca processes involving nucleons and hyperons. Over the past three decades, significant theoretical effort has been dedicated to understanding the thermal evolution of neutron stars with the nucleonic and hyperonic Urca processes emerging as dominant cooling mechanisms. While recent studies have proposed the presence of $\Delta$ isobars in the dense core of neutron stars, the specific contribution of the $\Delta$-direct-Urca process to neutrino emissivity has not been explicitly quantified. In this work, we address this gap by deriving the neutrino emissivity associated with $\Delta$-direct-Urca processes in dense matter inside neutron stars. We show that once $\Delta$-direct-Urca processes become kinematically allowed, their emissivities are comparable to those of nucleonic-direct-Urca processes.
\end{abstract}

\maketitle

\textit{Introduction--}   Recent astrophysical observations \cite{Demorest2010,vanKerkwijk2011,Antoniadis:2013pzd,Cromartie2020,Abbott2020} have established that neutron star (NS) masses may exceed 2 $M_\odot$. For these massive stars, the matter density inside the  core of the NS may reach several times the nuclear saturation density ($n_0$). At such high densities, additional exotic degrees of freedom may emerge beyond the nucleonic components. These exotic degrees of freedom include strange baryons \cite{Glendenning1982,Balberg1997}, nonstrange $\Delta$ isobars \cite{Oliveira:2007zz,Drago2014,JiaJie2018,Vishal2025}, boson condensates \cite{Haensel1981,Glendenning1998kaon,Parmar2024}, and deconfined strange quark matter \cite{Shapiro1983,Glendenning1985,Kumar2023}. Among these, $\Delta$ isobars are particularly relevant because they may soften the equation of state (EOS) at intermediate densities while restoring stiffness at higher densities, allowing $\Delta$-admixed EOSs to satisfy complementary constraints from gravitational-wave and NICER observations \cite{JiaJie2018}. However, the $\Delta$ optical potential in nuclear matter is poorly constrained, as the information on the $\Delta$-nucleon interaction is obtained only indirectly from nuclear processes \cite{Bodek2020, OConnell1990}. Consequently, uncertainties in its isoscalar and isovector components lead to considerable variations in the predicted onset density and abundance of $\Delta$ isobars in NS matter \cite{Drago2014,Kolomeitsev2017,Ribes2019}.

Theoretical models predict that, depending on the meson-$\Delta$ couplings inferred from the $\Delta$ optical potential, the $\Delta^-$ may appear below, near, or above the $\Lambda$-hyperon threshold, underscoring the importance of investigating its presence in NS matter \cite{Oliveira:2007zz,Drago2014,JiaJie2018,Dexheimer2021}.
The appearance of $\Delta^-$ isobars has several implications for NS properties. Because the $\Delta^-$ is negatively charged, it suppresses the abundance of negatively charged electrons ($e^-$) while enhancing that of positively charged protons ($p$). This, in turn, lowers the threshold density for the onset of the nucleonic-direct-Urca (NdUrca) process. Additionally, it opens up the possibility of semileptonic decays of $\Delta$ isobars \cite{Prakash1992,Yakovlev2001}.

These semileptonic processes inside NSs are important mechanisms of its cooling. NSs are born with extremely high internal temperatures, reaching 
$T \sim 10^{11}\,\mathrm{K}$ immediately after their birth in core-collapse supernovae \cite{Potekhin2015}. Owing to neutrino emitting processes operating in the dense interior, the temperature decreases within the first few seconds to $T \sim 10^{8}-10^{9}$ K. During the subsequent evolution, neutrino emission continues to dominate the energy loss for approximately $10^{5}$ yr. Below $T \sim 10^{6}\,\mathrm{K}$, the neutrino luminosity becomes
subdominant, and surface photon emission governs the thermal evolution
\cite{Yakovlev1999,Yakovlev2004}. The most efficient neutrino emitting process is NdUrca, $n\rightarrow p+e^-+\bar{\nu_e}$. However, this process is forbidden in the low density regime due to momentum triangular condition \cite{Lattimer1991,PHaensel1994},

\begin{equation}
|k_{F,B_2}-k_{F,l}| \leq k_{F,B_{1}} \leq k_{F,B_2}+k_{F,l},
\end{equation} where $k_{F,B_1}$, $k_{F,B_2}$ and $k_{F,l}$ denote the Fermi momenta of the participating baryons and the lepton, respectively. At such densities, the comparatively slower modified Urca process becomes operative and serves as the dominant neutrino-emission mechanism.

It has been shown that the early appearance of $\Delta^-$ isobars at densities 
$\lesssim 2n_0$ can increase the proton fraction sufficiently to  trigger the onset of the NdUrca process, which might otherwise remain kinematically forbidden in this density regime \cite{JiaJie2018}. 
Additionally, the presence of $\Delta$ isobars opens several new neutrino and antineutrino-emitting channels. 
As demonstrated in Ref.~\cite{Raduta2021}, the processes
$\Delta^- \rightarrow n + e^- + \bar{\nu}_e$ and
$\Delta^0 \rightarrow p + e^- + \bar{\nu}_e$, together with their
corresponding neutrino-emitting processes,
$n + e^- \rightarrow \Delta^- + \nu_e$ and
$p + e^- \rightarrow \Delta^0 + \nu_e$, respectively,
become kinematically allowed once the corresponding triangle
conditions among the Fermi momenta are satisfied.
Furthermore, Refs.~\cite{JiaJie2018,Prakash1992} show that another
weak decay channel,
$\Delta^- \rightarrow \Lambda + e^- + \bar{\nu}_e$,
together with its corresponding neutrino-emitting process
$\Lambda + e^- \rightarrow \Delta^- + \nu_e$,
can also operate in the same density regime.

If the presence of $\Delta$ isobars allows for the aforementioned neutrino emitting processes to be operative, then they can significantly influence the neutrino cooling processes and NS thermal evolution. In particular, changes in the particle fractions can shift the NdUrca threshold, while semileptonic $\Delta$ decays provide additional cooling channels. Calculating the emissivities of these processes is therefore essential for determining whether $\Delta$-admixed cores are compatible with observed NS surface temperatures and for constraining $\Delta$ interactions and the dense-matter EOS. In this Letter, we derive and calculate the neutrino emissivity associated with $\Delta$-direct-Urca (dUrca) processes, which become relevant with the appearance of $\Delta$ isobars. The resulting emissivity can be used to improve NS cooling calculations and simulations in future studies.

\textit{The neutrino emitting processes--}
The seminal works of \cite{ Prakash1992,Yakovlev1999} and recent studies \cite{JiaJie2018,Raduta2021,Vishal2025} have shown that, in the  presence of $\Delta$ isobars, not only can the NdUrca process operate at relatively low densities but also many efficient neutrino emitting $\Delta$-dUrca processes open up inside the NS. These are

\begin{eqnarray}
    \Delta^- &\rightarrow& n + e^- + \bar{\nu}_e,
    \qquad
    n + e^- \rightarrow \Delta^- + \nu_e\\[4pt]
    \Delta^- &\rightarrow& \Lambda + e^- + \bar{\nu}_e,
    \qquad
    \Lambda + e^- \rightarrow \Delta^- + \nu_e\\[4pt]
    \Delta^0 &\rightarrow& p + e^- + \bar{\nu}_e,
    \qquad
    p + e^-\rightarrow \Delta^0 + \nu_e.
\end{eqnarray}

Within the relativistic mean-field (RMF) framework, Ref.~\cite{Raduta2021} showed that, for suitable choices of coupling parameter space and $\Delta$ potential, the process $\Delta^- \rightarrow n + e^- + \bar \nu_e$ and its corresponding neutrino-emitting process can operate at densities of $(2--3) n_0$. Furthermore, Ref. \cite{JiaJie2018} demonstrated that for DD-ME2 and PKO3 parameterization the $\Delta^-\rightarrow\Lambda+e^-+\bar{\nu_e}$ process and its corresponding neutrino-emitting process can be operative shortly after the onset of $\Lambda$ hyperons for stars with masses $M\ge1.5~M_\odot$. Since the $\Delta^0$ appears only at relatively high densities, the process $\Delta^0\rightarrow p+e^-+\bar{\nu}_e$ and its corresponding neutrino-emitting process become operative at densities $\gtrsim 3n_0$, which lies below the central density of the maximum-mass NS for the  parameterization used in these works.

\textit{Emissivity of $\Delta$-dUrca processes--}
We now proceed to find the neutrino emissivity for the above mentioned $\Delta$-dUrca processes. The neutrino emissivity of a given weak  process is calculated by evaluating the energy carried away by neutrinos per unit volume and per unit time.

The neutrino emissivity calculations are performed assuming beta equilibrium and charge neutrality in dense NS matter. Accordingly, the relevant chemical equilibrium conditions are given by 
$\mu_n = \mu_p + \mu_e$, 
$\mu_{\Delta^-} = \mu_n + \mu_e$, 
$\mu_{\Delta^0} = \mu_n$, 
and 
$\mu_{\Lambda} = \mu_n$, 
 where $\mu_i$ denotes the chemical potential of particle species $i$.

The three processes mentioned in the previous section can be generally written as
\begin{equation}
\Delta(k_\Delta,s_\Delta)
\;\rightarrow\;
B(k_B,s_B)
+
e^-(k_e)
+
\bar{\nu}_e(k_\nu),
\end{equation}
where $\Delta$ generically denotes the $\Delta^-$ and $\Delta^0$ states, and $B$ denotes $p$, $n$ or $\Lambda$.  

The general expression for neutrino emissivity is given by \cite{Yakovlev2001}
 \begin{equation}
Q = 4 \int 
\frac{d^3 k_\Delta}{(2\pi)^3}
\frac{d^3 k_B}{(2\pi)^3}
\frac{d^3 k_e}{(2\pi)^3}
\frac{d^3 k_{\nu}}{(2\pi)^3}
W_{fi}
E_\nu
f_\Delta(1 - f_B)(1 - f_e)
\end{equation}
where the prefactor 4 accounts for the $\Delta$ isobar spin, $f_i$'s are the Fermi-Dirac distribution function for the $i$-th species and the decay width is 
\begin{eqnarray}
    W_{fi}= (2\pi)^4 \delta (E_\Delta-E_B-E_e-E_\nu) \nn\\
    \times\delta(k_\Delta-k_B-k_e-k_\nu) \sum_{spin}|\mathcal{M}|^2.
    \label{eq:Wfi}
\end{eqnarray}
Here, $\mathcal{M}$ is the matrix element between the initial and final states.

The nucleonic and hyperonic dUrca processes are semileptonic weak decays of spin-$1/2$ baryons, in which a baryon transitions to another spin-$1/2$ baryon accompanied by the emission of a charged lepton and an antineutrino via the charged weak current \cite{Prakash1992,Yakovlev1999,Yakovlev2001}. In contrast, the $\Delta$ decay considered here involves a transition from a spin-$3/2$ baryon to a spin-$1/2$ baryon  accompanied by the emission of a charged lepton and an antineutrino.  The weak interaction Hamiltonian responsible for this transition is given by \cite{Wang2022}
\begin{equation}
    \mathcal{H}_i = \frac{g^2}{2}\{\bar \psi_B \Gamma^{\mu\sigma}W_\mu\psi_{\Delta,\sigma}\}\{\bar \psi_e \gamma_\nu(1-\gamma_5)W_\nu \ \psi_{\bar\nu}\},
    \nn
\end{equation}
where $g$ is the weak coupling constant and $W_\mu$ is the charged weak boson field. The hadronic transition vertex $\Gamma^{\mu\sigma}$ encodes the spin-$3/2 \to 1/2$ structure of the weak current and is written as \cite{Wang2022,Zhao2024}
\begin{equation}
     \Gamma^{\mu\sigma} = \gamma^{\mu}\Big[\frac{p^{\sigma}_{B}}{m_{B}}(f_1+g_1\,\gamma_5)\Big],
\end{equation}
with $f_1$ and $g_1$ being the vector and axial weak form factors. 

From this expression, the matrix element of this weak decay can be written as 
\begin{eqnarray}
    \mathcal{M}= \frac{G_F}{\sqrt{2}}C\{\bar u_B (k_B)\Gamma^{\mu\sigma}u_{\Delta\sigma}(k_\Delta)\} \nn\\
    \times\{\bar u_e(k_e) \gamma_\mu(1-\gamma_5) \ u_{\bar\nu} (k_\nu)\} ,
\end{eqnarray}
where $C$ is the Cabibbo factor, $G_F=\frac{g^2}{2M_W^2}$ is the Fermi constant, and $M_W$ is the $W$-Boson mass. The hadronic current involves a Rarita–Schwinger spinor $u_{\Delta,\sigma}$ for the spin-$3/2$ particle, while the leptonic current has the standard V–A structure. 
For the emissivity calculation, one proceeds by squaring the matrix element, summing over final spins and averaging over initial spins. 
\begin{equation}
      \sum_{spin} |\mathcal{M}|^2 =  \sum_{spin}\mathcal{M}^\dagger \mathcal{M}= C^2\frac{G_F^2}{2}H^{\mu\nu}L_{\mu\nu} \\
       \label{M^2}
\end{equation}
where the  leptonic tensor $L_{\mu\nu}$ and the hadronic tensor $H^{\mu\nu}$ can be calculated using Clifford algebra identities. The standard leptonic tensor is given by
\cite{Yakovlev2001}

\begin{eqnarray}
 L_{\mu\nu} =  \mathrm{Tr} \Big[ \slashed{p}_e \, \gamma_\mu (1-\gamma_5) \, \slashed{p}_\nu \, \gamma_\nu (1-\gamma_5) \Big]\nn \\ 
= 8 \Big( p_{e,\mu} p_{\nu,\nu} + p_{e,\nu} p_{\nu,\mu} - g_{\mu\nu} \, (p_e \cdot p_\nu) \nn\\
- i \, \epsilon_{\mu\nu\alpha\beta}\, p_{e\alpha} p_{\nu\beta} \Big)
\label{eq:Lmunu}  
\end{eqnarray}
Unlike ordinary Urca processes, in which the initial baryon has spin $1/2$, the present process involves a spin-$3/2$ $\Delta$ resonance described by a Rarita--Schwinger spinor. Consequently, the summation over its spin states produces a nontrivial projection operator rather than the simple Dirac structure. The hadronic tensor is therefore constructed using the Rarita--Schwinger projector \cite{Rarita1941,Lovas1998}. In particular, the spin-$3/2$ completeness relation is given by \cite{Najjar2025},
\begin{eqnarray}
P^{(3/2)}_{\alpha\beta} &=& -(\slashed p + m)
\left(g^{\alpha\beta} -\frac{1}{3} \gamma^{\alpha}  {\gamma}^{\beta} \right. \nonumber \\
&& \left. - \frac{2 p^{\alpha} p^{\beta}}{3 m^2} 
+ \frac{{ p^{\alpha}\gamma}^{\beta} -  p^{\beta}{\gamma}^{\alpha}}{3 m}\right).
\label{eq:projection operator}
\end{eqnarray}
The evaluation of the hadronic tensor involves lengthy Dirac algebra due to the presence of the spin-$3/2$ projection operator. 
To perform these calculations, we employed the FeynCalc package in Mathematica \cite{Shtabovenko2020}, which allows for automated symbolic computation of Dirac traces and tensor contractions.

With the above completeness relation, the hadronic tensor takes the form,
\begin{equation}
\begin{aligned}
H^{\mu\nu}
={}&
\mathrm{Tr}\Big[
(\slashed{p}_B+m_B)
\Gamma^{\mu\sigma}
(\slashed{p}_\Delta+m_\Delta)
P^{(3/2)}_{\sigma\rho}
\Gamma^{\nu\rho}
\Big]
\\
={}&
\frac{2\left[
m_\Delta^2p_B^2
-(p_\Delta\!\cdot p_B)^2
\right]}
{m_\Delta^2m_B^2}
\Bigg\{
(f_1^2+g_1^2)
\\
&\times
\Big[
-g^{\mu\nu}(p_\Delta\!\cdot p_B)
+p_\Delta^\nu p_B^\mu
+p_\Delta^\mu p_B^\nu
\Big]
\\
&
+(f_1^2-g_1^2)
m_\Delta m_B g^{\mu\nu}
-2if_1g_1\epsilon^{\mu\nu\alpha\beta}
p_{\Delta\alpha}p_{B\beta}
\Bigg\}.
\end{aligned}
\end{equation}

Consequently, the desired squared matrix element can be written as,
\begin{eqnarray}
\small
    \sum_{spin}|\mathcal{M}|^2=&\frac{4G_F^2C^2}{3\,m_\Delta^2 m_B^2}\Big[(m_\Delta^2-m_B^2)^2-2m_e^2(m_\Delta^2+m_B^2)\Big] \nn\\
   & \times \Big[(f_1^2+3g_1^2)\,+(f_1^2-g_1^2).
v_e.v_\nu\Big]
\label{eq:M^2}
\end{eqnarray}

The evaluation of the phase-space integrals can be substantially simplified by employing the phase-space decomposition technique, which is particularly effective for computing process rates in strongly degenerate matter \cite{Shapiro1983,Yakovlev2001}. 
In the present case, the phase-space integral is dominated by thin regions of momentum space in the immediate vicinity of the participating particles’ Fermi surfaces, reflecting the strong degeneracy of these constituents. As a result,
the particle momenta may be set equal to their respective Fermi momenta in all smooth
functions appearing under the integral.
In the degenerate limit, the momentum integrals may be written in terms of
densities of states near the Fermi surfaces \cite{Yakovlev2001}.

\begin{equation}
\frac{d^3 k_\Delta}{(2\pi)^3}\,
\frac{d^3 k_p}{(2\pi)^3}\,
\frac{d^3 k_e}{(2\pi)^3}
\;\longrightarrow\;
\prod_{i=\Delta,B,e}
\frac{k_{F_i} m_i^\ast}{(2\pi)^3}\,
dE_i\, d\Omega_i
\label{Eq:phase-space1}
\end{equation}
where $d\Omega_i$ is the solid angle element and $m_i^* = p_{F_i}/v_{F_i}$ is the effective mass, with $v_{F_i}$ the Fermi velocity. For ultra relativistic neutrinos,
\begin{equation}
    \frac{d^3 k_\nu}{(2\pi)^3}\longrightarrow \frac{E_\nu^2 dE_\nu d\Omega_\nu}{(2\pi)^3}.
    \label{Eq:phase-space2}
\end{equation}

Substituting Eqs.~(\ref{eq:Wfi})--(\ref{Eq:phase-space2}) into the general expression and applying the standard integration techniques of Ref.~\cite{Yakovlev2001}, we obtain
\begin{align}
Q &= \frac{457\pi G_F^2 C^2}{3780\, m_\Delta^2 m_B^2}
\, T_9^6
\Big[(m_\Delta^2 - m_B^2)^2 
- 2 m_e^2 (m_\Delta^2 + m_B^2)\Big] \nonumber \\
&\quad \times \left(f_1^2 + 3 g_1^2\right)
\, m^*_\Delta m^*_B m^*_e
\, \Theta_{\Delta B e},
\label{eq:Q}
\end{align}
where $m_e^\ast=\mu_e/c^2$, $\mu_e$ is the electron chemical potential,
$c$ is the speed of light in vacuum, and $T_9=T/(10^9~\mathrm{K})$ denotes
the temperature in units of $10^9$ K. The quantity $\Theta_{\Delta B e}$ is the threshold function that enforces the kinematic threshold of the corresponding dUrca process. The emissivity is derived under the low-temperature, strongly degenerate
approximation, $k_B T \ll E_{F,i}$, where $k_B$ is the Boltzmann constant
and $E_{F,i}$ denotes the Fermi energy of the $i$-th particle species.
The final expression includes the contribution from both the direct decay process $\Delta \rightarrow B + e^- + \bar{\nu}_e$ and the corresponding neutrino-emitting process $B + e^- \rightarrow \Delta + \nu_e$, which contribute equally under chemical equilibrium conditions.

\textit{Implications for cooling--}
To illustrate the early onset of the NdUrca process and the opening of additional $\Delta$ isobar dUrca channels, we consider the matter within the density dependent RMF model and take DD-ME2 parameterization \cite{Lalazissis2005,Vivek2021}. We consider the matter is composed of nucleons, hyperons and $\Delta$ isobars. We adopt $x_{\sigma\Delta}=1.18$, $x_{\omega\Delta}=1.0$, and $x_{\rho\Delta}=1.0$. The $\Delta$-meson coupling ratios $x_{\sigma\Delta}$, $x_{\omega\Delta}$, and $x_{\rho\Delta}$ denote the ratios of the $\Delta$-meson coupling constants to the corresponding nucleon-meson coupling constants for the $\sigma$, $\omega$, and $\rho$ mesons, respectively. 
A detailed comparison of different EOS models and $\Delta$-meson coupling schemes is presented in Supplemental Material, demonstrating that the $\Delta$-dUrca processes remain operative over a broad region of parameter space \cite{SM}.

To evaluate numerically $\Delta$-dUrca emissivity using eq. \eqref{eq:Q} we consider $G_F=1.436 \times 10^{-49}$ erg cm$^3$, $C = 0.974 $ for strangeness conserving process and $C = 0.231$ for strangeness changing process. We adopt $T=10^9~\mathrm{K}$ corresponding to
$k_B T \simeq 0.086~\mathrm{MeV}$,
which is much smaller than the characteristic Fermi energies of the particles participating in the Urca processes. The vector ($f_1$) and axial-vector ($g_1$) coupling constants in Eq. \eqref{eq:Q}  are well established for the standard NdUrca process, with typical values $f_1 = 1$ and $g_1 \simeq 1.26$ \cite{Yakovlev2001,Dominguez2023} , derived from weak interaction phenomenology. Similarly, for hyperonic dUrca processes, these couplings are reasonably well constrained based on experimental inputs from hyperon beta decays \cite{Prakash1992,Gaillard:1984ny,Cabibbo2003}. However, in the case of $\Delta$-dUrca processes, the corresponding $f_1$ and $g_1$ are not well determined due to the lack of experimental constraints on $\Delta$ isobar weak interactions. 

Therefore, in the present work, we consider a range of values for the vector and axial-vector couplings, $f_1 \in [-2,2]$ and $g_1 \in [-1.5,\,1.5]$, to explore their impact on the neutrino emissivity. The chosen range encompasses the typical values adopted for both nucleonic and hyperonic dUrca processes \cite{Prakash1992}, ensuring that our analysis remains physically consistent. For the chosen range of $f_1$ and $g_1$, the emissivity varies from a
minimum value of $\sim 4.60\times10^{23}$ to a maximum value of
$\sim 6.60\times10^{24}~\mathrm{erg\,cm^{-3}\,s^{-1}}$.

\begin{figure}[t]
    \centering
    \includegraphics[width=0.875\linewidth]{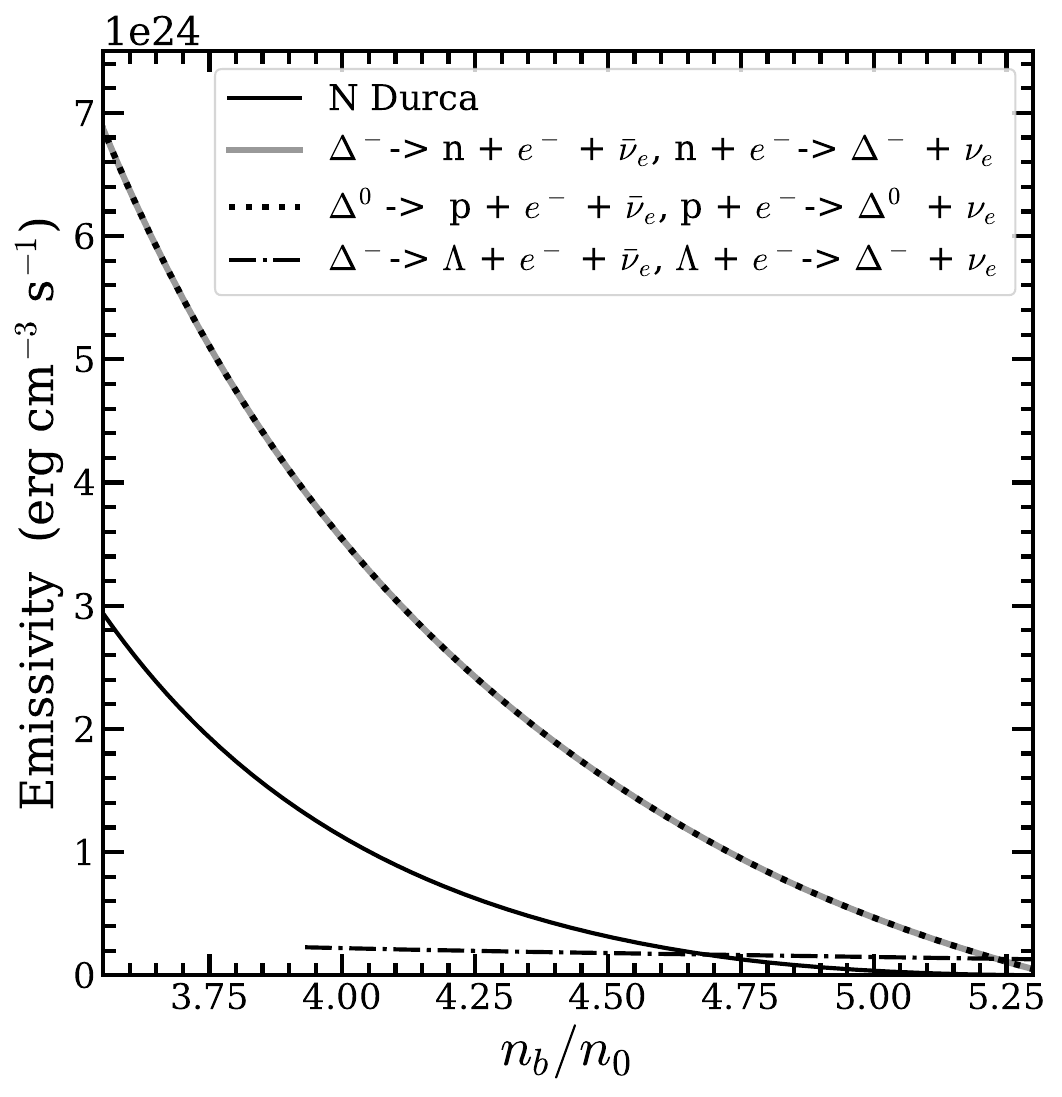}
    \caption{Emissivities of the NdUrca and $\Delta$-dUrca processes in a $1.7M_\odot$ NS.}
    \label{fig:emissivity_comparison}
\end{figure}

 For consistency and in the absence of precise constraints, we adopt the same vector and axial-vector coupling constants used for NdUrca in $\Delta$-dUrca emissivity calculations. For the strangeness-changing process $\Delta^- \rightarrow \Lambda + e^- + \bar{\nu}_e$ and its corresponding neutrino-emitting process, we employ hyperon weak coupling values, taking $f_1=-1.225$ and $g_1=0.893$ \cite{Yakovlev2000}. With these values of $f_1$ and $g_1$ we compare the $\Delta$-dUrca emissivities with the NdUrca emissivity in Fig. \ref{fig:emissivity_comparison}. The emissivities of the NdUrca process and the $\Delta$-dUrca processes, $\Delta^-\rightarrow n+e^-+\bar{\nu}_e$, $\Delta^0\rightarrow p+e^-+\bar{\nu}_e$, and $\Delta^-\rightarrow\Lambda+e^-+\bar{\nu}_e$, including the contributions from their corresponding neutrino-emitting processes are plotted as a function of matter density. We note that the emissivities gradually decrease with increasing baryon density. This behavior originates from the density dependence of the effective masses, chemical potentials, and particle composition.

It is evident that the $\Delta$-dUrca processes $\Delta^-\rightarrow n + e^- +\bar\nu_e$ and $\Delta^0\rightarrow p + e^- +\bar\nu_e$ and their corresponding neutrino-emitting processes yield  neutrino emissivities that are slightly higher than that of the NdUrca process over the entire density range considered. However, the difference between the two remains small, indicating that both processes operate with nearly comparable efficiency. This demonstrates that the $\Delta$-dUrca processes  are as efficient as the standard NdUrca process and can therefore act as an equally powerful mechanism for rapid neutrino cooling in NS interiors.

This behavior highlights an important physical implication: once the $\Delta$-dUrca channel is kinematically allowed, it can contribute significantly to the neutrino cooling of NSs. Although slightly stronger than the NdUrca process, the $\Delta$-dUrca process remains a fast cooling mechanism and therefore cannot be neglected. We expect this conclusion to be largely independent of the chosen parameterization of the matter properties.

Therefore, the inclusion of $\Delta$ isobars in the EOS not only lowers the NdUrca threshold,  but also introduces additional rapid cooling pathways by opening the channels of $\Delta$-dUrca. Overall, the results suggest that if the $\Delta$-dUrca processes $\Delta^-\rightarrow n + e^- +\bar\nu_e$, $\Delta^0\rightarrow p + e^- +\bar\nu_e$ and their corresponding neutrino-emitting processes are allowed,  their emissivities are comparable to that of the NdUrca process, making it a crucial ingredient in realistic NS cooling simulations.

\textit{Conclusions--}
The dUrca processes constitute one of the most efficient mechanisms governing the thermal evolution of NSs. 
In particular, hyperonic dUrca channels are known to accelerate cooling relative to modified Urca processes, and their emissivities have been extensively studied in the literature ~\cite{Lattimer1991,Prakash1992,Raduta2018,Raduta2019,Yakovlev2001}. Despite indications that $\Delta$ isobars may appear in dense matter at supranuclear densities and potentially participate in weak interaction processes, a quantitative description of their contribution to neutrino emissivity has so far been lacking. 

In this work, for the first time, we derive a general expression for the neutrino emissivity associated with $\Delta$-dUrca processes, explicitly incorporating the relevant weak interaction matrix elements involving the Rarita–Schwinger spinor.
 
 We employ the DD-ME2 parameterization within the RMF framework to determine the kinematically allowed regions of the NdUrca and $\Delta$-dUrca processes in NSs containing $\Delta$ isobars. We then calculate the neutrino emissivity of $\Delta$-dUrca processes and compare them with the NdUrca emissivity. We find that $\Delta$-dUrca emissivities of $\Delta^-\rightarrow n + e^- +\bar\nu_e$, $\Delta^0\rightarrow p + e^- +\bar\nu_e$ and their corresponding neutrino-emitting processes are slightly higher than that of NdUrca over the relevant density range. This indicates that, once kinematically allowed, $\Delta$-dUrca processes act as equally efficient fast cooling mechanisms. Consequently, the presence of $\Delta$ isobars introduces additional rapid cooling channels, which can have significant implications for the thermal evolution of massive NSs. These findings highlight the necessity of incorporating $\Delta$ isobars in realistic NS cooling simulations.

\textit{Acknowledgments--}
The authors acknowledge financial support from the Science
and Engineering Research Board (SERB), Department of Science and Technology, Government of India through Project No. CRG/2022/000069.
\nocite{Niksic2002,Lalazissis2005,Typel2005,Typel2010,TANINAH2020,Abbott2018,Riley:2019yda,Riley:2021pdl,Miller:2019cac,Miller:2021qha,Romani:2022jhd}
\bibliography{references}
\clearpage
\onecolumngrid
\section{Supplemental material}

To examine the model dependence of our results, we consider five density-dependent relativistic mean-field EOSs: DDME2  \cite{Lalazissis2005}, DDME1 \cite{Niksic2002}, DDMEX \cite{TANINAH2020}, DD2 \cite{Typel2010}, and DD1 \cite{Typel2005}. 
These models span a representative range of nuclear saturation properties, symmetry-energy behavior, and high-density stiffness. Table~\ref{tab:eosparams} summarizes the corresponding saturation properties, $-E_0$ the binding energy per nucleon, $K_0$ the incompressibility coefficient, $E_{\rm sym}$ the symmetry energy, $L_{\rm sym}$ the slope parameter,  $K_{\rm sym}$ the curvature parameter of the symmetry energy at saturation density and $1.4$$M _\odot$ properties such as ,$R_{1.4} $ radius and $\Lambda_{1.4}$, dimensionless tidal deformability \cite{Vivek2021}. All models reasonably satisfy present observational constraints of gravitational wave observations and massive neutron stars.

\begin{table*}[ht]
\caption{Nuclear matter saturation properties, selected neutron-star properties and maximum mass for the density-dependent RMF EOSs used in this work. }
\label{tab:eosparams}
\centering
\begin{tabular}{lccccccccc}
\hline\hline
\\[-1.8ex]
EOS &
$n_0$ &
$-E_0$ &
$K_0$ &
$E_{\rm sym}$ &
$L_{\rm sym}$ &
$K_{\rm sym}$ &
$R_{1.4}$ &
$\Lambda_{1.4}$ &
$M_{\max}$ \\
&
(fm$^{-3}$) &
(MeV) &
(MeV) &
(MeV) &
(MeV) &
(MeV) &
(km) & &
($M_\odot$) \\[0.4ex]
\hline
\\[-1.6ex]

DD1
& 0.1487
& 16.021
& 240.00
& 31.60
& 55.949
& $-95.24$
& 13.126
& 678.92
& 2.410
\\

DD2
& 0.1490
& 16.020
& 242.70
& 32.73
& 54.966
& $-93.24$
& 13.133
& 683.97
& 2.418
\\

DD-ME1
& 0.1520
& 16.200
& 244.50
& 33.10
& 55.370
& $-101.07$
& 13.086
& 672.37
& 2.443
\\

DD-ME2
& 0.1520
& 16.140
& 250.89
& 32.30
& 51.253
& $-87.31$
& 13.146
& 706.08
& 2.483
\\

DD-MEX
& 0.1520
& 16.140
& 267.059
& 32.269
& 49.576
& $-71.47$
& 13.293
& 773.49
& 2.556
\\
\hline\hline
\end{tabular}
\end{table*}

\begin{table}[htbp]
\centering
\caption{Direct Urca onset densities in units of saturation density $\rho_0$ for different  models, evaluated for the coupling choice $x_{\sigma\Delta}=1.18$, $x_{\omega\Delta}=1.0$, and $x_{\rho\Delta}=1.0$. Results are shown for hyperonic matter ($npY$) and hyperonic matter including $\Delta$ baryons ($npY\Delta$). A dash indicates that the corresponding process does not occur.}
\label{tab:durca_onsets}
\begin{tabular}{lcccccc}
\hline\hline
EOS & Composition &
$\rho_{N}$ &
$\rho_{\Delta^- n}$ &
$\rho_{\Delta^0 p}$ &
$\rho_{\Delta^- \Lambda}$ \\
\hline

DD1
 & $npY$          & --    & --    & --    & --    \\
 & $npY\Delta$    & 1.84  & 3.45  & 3.66  & 4.15  \\

DD2
 & $npY$          & --    & --    & --    & --    \\
 & $npY\Delta$    & 1.84  & 3.42  & 3.63  & 4.13  \\

DD-ME1
 & $npY$          & --    & --    & --    & --   \\
 & $npY\Delta$    & 1.80  & 3.36  & 3.56  & 3.99  \\

DD-ME2
 & $npY$          & --    & --    & --    & --  \\
 & $npY\Delta$    & 1.85  & 3.35  & 3.55  & 3.95  \\

DD-MEX
 & $npY$          & --    & --    & --    & --   \\
 & $npY\Delta$    & 1.79  & 3.21  & 3.39  & 3.74  \\

\hline\hline
\end{tabular}
\end{table}
\newpage

\begin{figure}[t]
    \centering
    \includegraphics[width=0.8\linewidth]{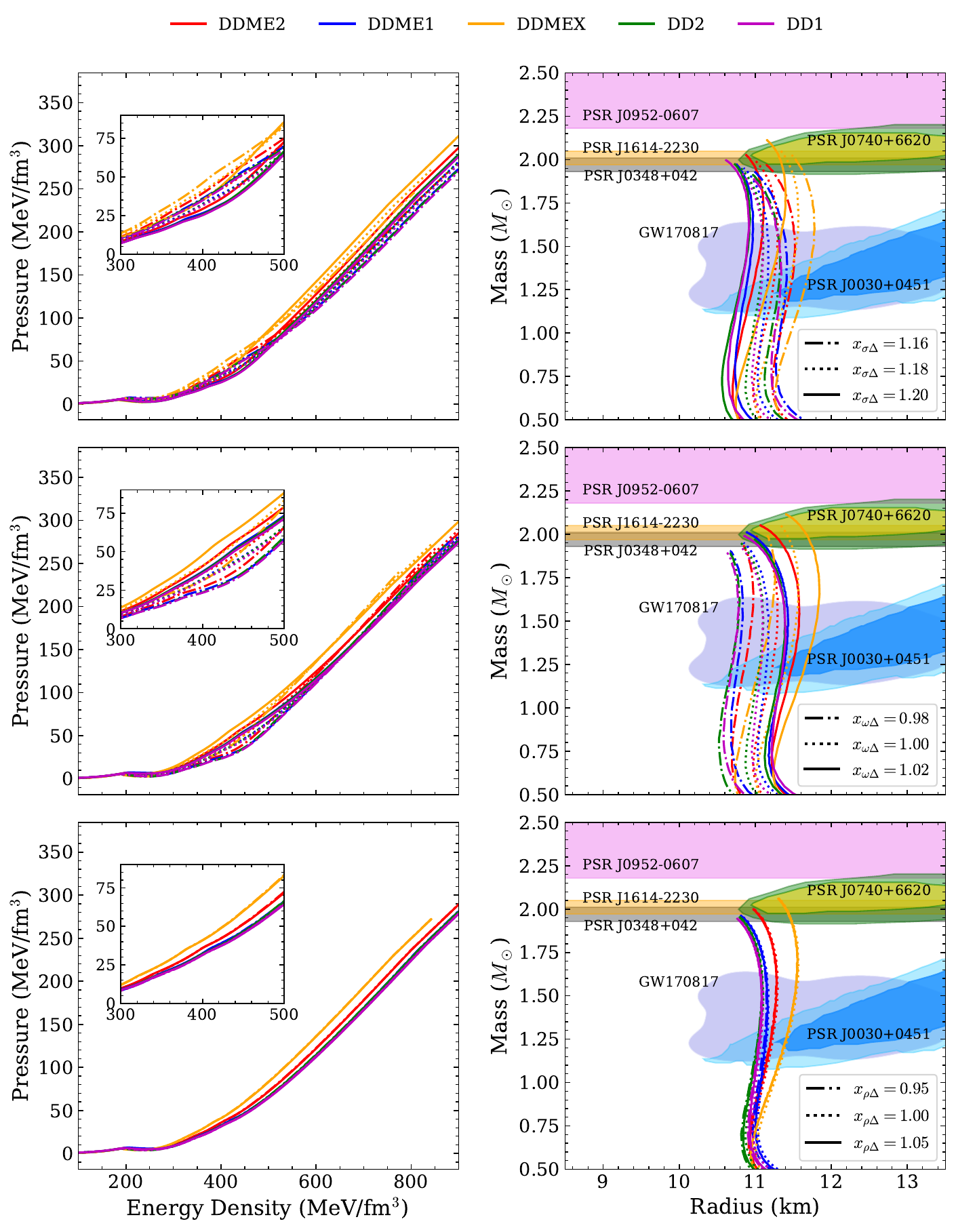}
    \caption{\textit{Left panel:} Variation of pressure with energy densities. \textit{Right panel: }Mass-radius curves for the used EoSs. Different curves for a particular EOS are for different $x_{\sigma\Delta}$ ({\it top panel:}), for different $x_{\omega\Delta}$ ({\it middle panel:}) and for different $x_{\rho\Delta}$ ({\it bottom panel:}). The mass and radius constraints are taken from \cite{Abbott2018,Riley:2019yda,Riley:2021pdl,Miller:2019cac,Miller:2021qha,Romani:2022jhd}.}
    \label{fig:EOS}
\end{figure}

 The pressure--energy density relations and mass--radius curves for the selected EOSs are shown in Fig.~\ref{fig:EOS}.  and all the models support neutron stars with massive neutron stars with masses  $ \gtrsim 2M_\odot$.
 The three rows correspond to representative variations of the $\Delta$-meson couplings $x_{\sigma\Delta}$, $x_{\omega\Delta}$, and $x_{\rho\Delta}$, respectively. The selected EOSs span a broad range of stiffness while remaining compatible with current observational constraints. 

In the presence of $\Delta$ baryons, the threshold density of nucleonic dURCA and $\Delta$ dURCA processes lower substantially. The comparison of threshold densities of different processes in the absence and presence of $\Delta$ baryons is shown in table \ref{tab:durca_onsets}. It is found that while in the absence of $\Delta$ baryons, the dURCA processes are forbidden, in the presence of $\Delta$ baryons, they are allowed after a reasonable threshold density.

\begin{figure}[t]
    \centering
    \includegraphics[width=0.6\linewidth]{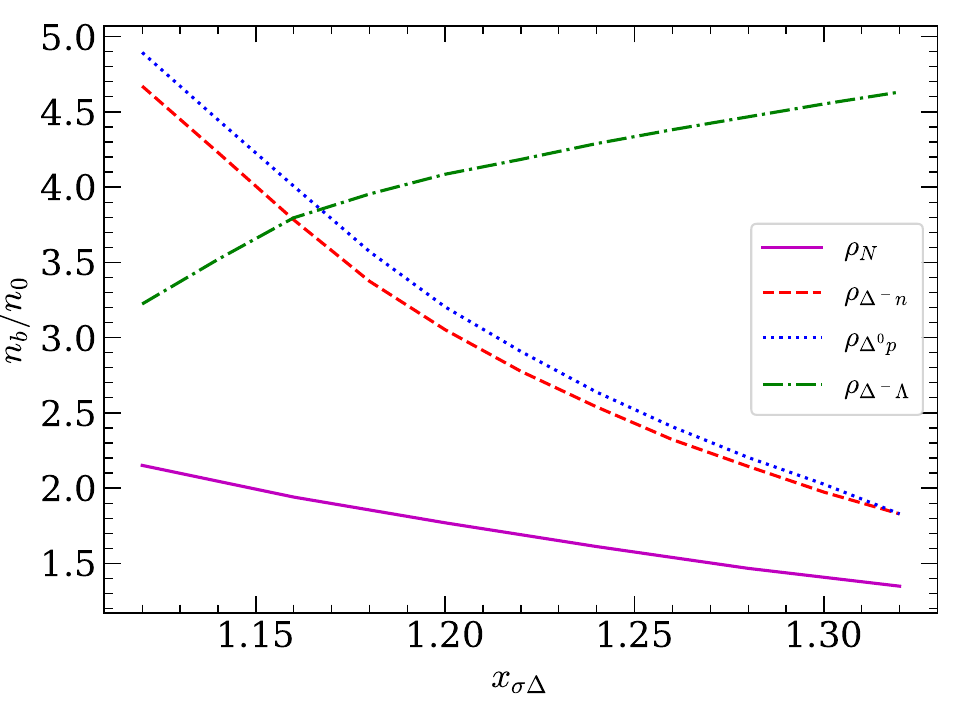}
    \caption{Onset densities of NdUrca, and $\Delta$ dUrca processes for different $x_{\sigma\Delta}$, while  $x_{\omega\Delta}$ and $x_{\rho\Delta}$ are fixed at 1.0 for DDME2 EOS.}
    \label{fig:onset_densities}
\end{figure}

We further explore a broad range of $x_{\sigma\Delta}$ coupling and present the corresponding onset densities of NdUrca and $\Delta$ dUrca processes in Fig.~\ref{fig:onset_densities}. 
 Increasing $x_{\sigma\Delta}$ enhances the attractive scalar interaction and leads to an earlier appearance of the $\Delta^-$ baryon in dense matter. Consequently, the threshold densities of the $\Delta$ dUrca channels shift to lower densities with increasing $x_{\sigma\Delta}$. The figure demonstrates that the $\Delta$-induced dUrca processes remain operative over a broad range of coupling strengths.

\begin{figure}[t]
    \centering
    \includegraphics[width=1.0\linewidth]{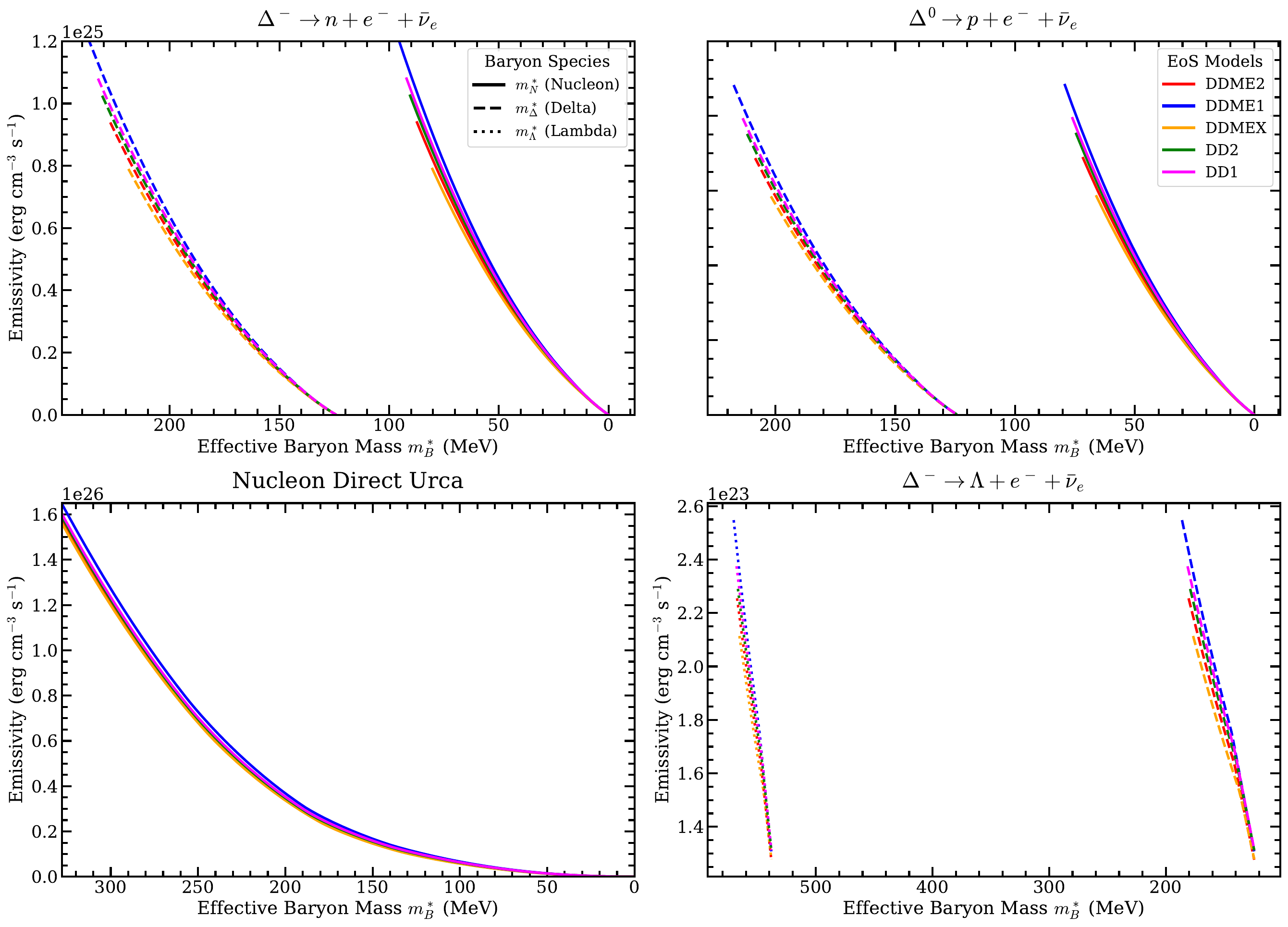}
    \caption{Variation of emissivities of NdUrca process and $\Delta$ dUrca processes with effective baryon masses, evaluated for coupling choice of $x_{\sigma\Delta}=1.18$, $x_{\omega\Delta}=1.0$, and $x_{\rho\Delta}=1.0$. }
    \label{fig:emissivities_vs_effectivemass}
\end{figure}
Dependence of the neutrino emissivities on the effective masses of the participating baryons for the discussed dURCA processes for different EOS models are shown in Fig.~\ref{fig:emissivities_vs_effectivemass}. The dashed and solid curves correspond to the variation with effective masses of the initial and final baryons, respectively. In all channels, the emissivity decreases with decreasing effective baryon mass, reflecting the explicit mass dependence of the phase-space factors in Eq.~(15). Quantitative differences among the EOSs arise from their different density dependence of the in-medium effective masses. The relatively stiffer EOS DDMEX generally yields lower emissivities, whereas the softer EOS DD1 produces comparatively higher emissivities.

\begin{figure}[t]
    \centering
    \includegraphics[width=1.0\linewidth]{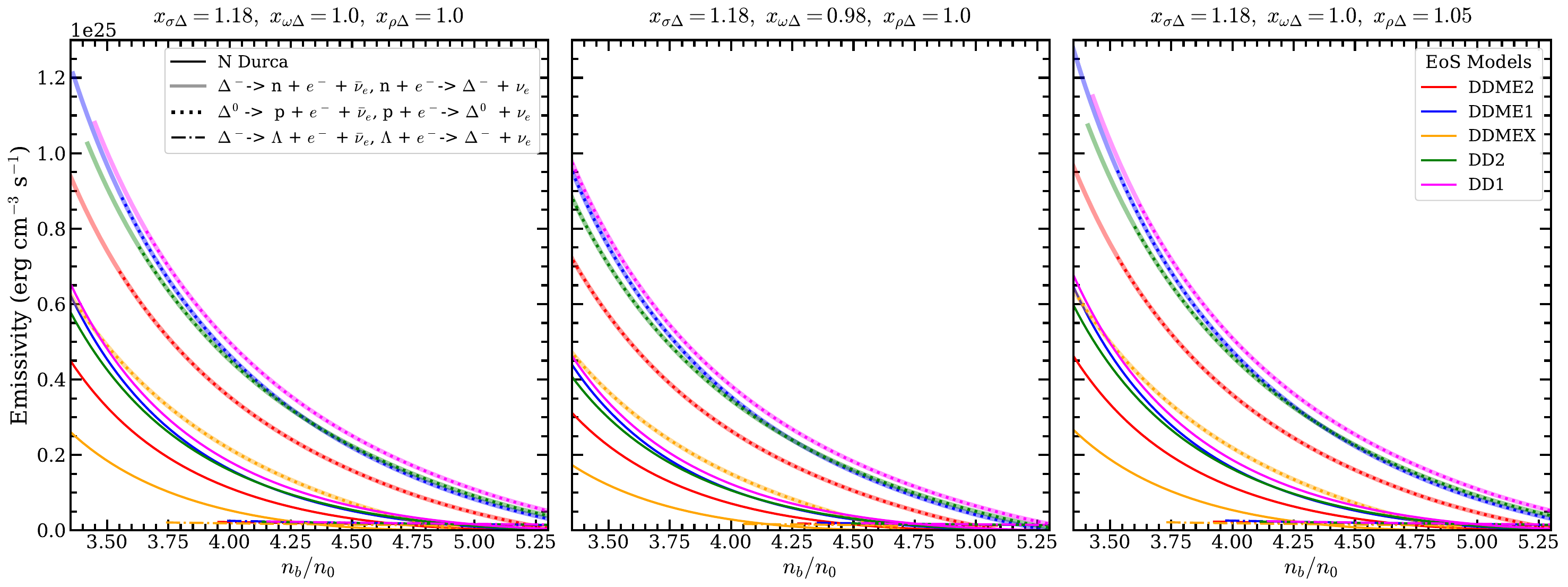}
    \caption{Variation of emissivities of NdUrca process and $\Delta$ dUrca processes with the baryon density, evaluated for different coupling choices of $x_{\sigma\Delta}$, $x_{\omega\Delta}$, and $x_{\rho\Delta}$ in different panels for DDME2 EOS. }
    \label{fig:emissivities_All_models}
\end{figure}
Density dependence of the neutrino emissivities of the NdUrca process and the $\Delta$-induced dUrca channels for the five EOS models considered in this work is shown in Fig.~\ref{fig:emissivities_All_models}. The three panels correspond to representative choices of the $\Delta$ couplings. Although quantitative differences among EOSs are present, the $\Delta$ dUrca channels remain comparable to or exceed the standard nucleonic dUrca emissivity over a broad density except for $\Delta^- \rightarrow \Lambda+e^- + \bar\nu_e$ process. The relatively stiffer EOS DDMEX generally yields lower emissivities, whereas the softer EOS DD1 produces systematically higher emissivities. This behavior reflects the sensitivity of the emissivity to the dense-matter composition and effective masses determined by the underlying EOS.
\nocite{Niksic2002,Lalazissis2005,Typel2005,Typel2010,TANINAH2020,Abbott2018,Riley:2019yda,Riley:2021pdl,Miller:2019cac,Miller:2021qha,Romani:2022jhd}

\end{document}